\documentclass[11pt,preprint]{aastex}

\usepackage{emulateapj5}
\usepackage{epsf}

\newcommand{\tx}[1]{\textrm{#1}}

\newcommand{\kms}{km~$\tx{s}^{-1}$}

\newcommand{\Io}{I$_{814}$}
\newcommand{\Vs}{V$_{606}$}

\newcommand{\Rekpc}{$R_{\tx{e}}$}

\newenvironment{inlinefigure}{
\def\@captype{figure}
\noindent\begin{minipage}{0.999\linewidth}\begin{center}}
{\end{center}\end{minipage}\smallskip}

\slugcomment{ApJ Letters, in press}

\shorttitle{The Fundamental Plane of field early-type galaxies}
\shortauthors{Treu et al.}

\begin{document}

\title{The evolution of field early-type galaxies to
$z\sim0.7$}\footnote{Based on observations collected at ESO (Paranal)
under programmes 65.O-0446 and 66.A-0362, and with the NASA/ESA HST,
obtained at STScI, which is operated
by AURA, under NASA contract NAS5-26555.}

\author{Tommaso Treu} \affil{California Institute of Technology,
Astronomy Department, 105-24, Pasadena, CA 91125
\email{tt@astro.caltech.edu}}

\and

\author{Massimo Stiavelli, S. Casertano} \affil{Space Telescope Science Institute, 3700
San Martin Dr., Baltimore, MD 21218}

\and

\author{Palle M{\o}ller} \affil{ European Southern Observatories,
Karl-Schwarzschild Str. 2, D85748, Garching bei M\"unchen, Germany}

\and

\author{Giuseppe Bertin} \affil {Universit\`a degli Studi di Milano,
Dipartimento di Fisica, Via Celoria 16, I20133, Milano, Italy}

\begin{abstract}
We have measured the Fundamental Plane (FP) parameters for a sample of
30 {\it field} early-type galaxies (E/S0) in the redshift range
$0.1<z<0.66$. We find that: {\it i)} the FP is defined and tight out
to the highest redshift bin; {\it ii)} the intercept $\gamma$ evolves
as $d\gamma/dz=0.58^{+0.09}_{-0.13}$ (for $\Omega=0.3,
\Omega_{\Lambda}=0.7$), or, in terms of average effective mass to
light ratio, as $d\log(M/L_B)/dz=-0.72^{+0.11}_{-0.16}$, i.~e.  faster
than is observed for cluster E/S0 ($-0.49\pm0.05$). In addition, we
detect [OII] emission $>5$\AA\, in 22\% of an enlarged sample of 42
{\it massive} E/S0 in the range $0.1<z<0.73$, in contrast with the
quiescent population observed in clusters at similar $z$. We interpret
these findings as evidence that a significant fraction of massive
field E/S0 experiences secondary episodes of star-formation at $z<1$.
\end{abstract}


\section{Introduction}

The Fundamental Plane (Djorgovski \& Davis 1987; Dressler et al.\
1987; hereafter FP) is a tight correlation between the effective
radius (R$_e$), the effective surface brightness (SB$_e$), and the
central velocity dispersion ($\sigma$)
\begin{equation}
\label{eq:FP} 
\log R_{\tx{e}} = \alpha \log~\sigma + \beta~SB_{\tx{e}} + \gamma 
\end{equation} 
that is observed to hold for local early-type galaxies (E/S0). Under
the assumption that E/S0 are homologous dynamical systems the FP can
be interpreted in terms of a power law relation between mass ($M$) and
mass-to-light ratio ($M/L$) (Treu et al.\ 2001b and references
therein). In recent years, it has been found that a tight FP exists in
clusters out to redshift $z=0.83$ (van Dokkum \& Franx 1996; Pahre
1998; Bender et al.\ 1998; Kelson et al.\ 2000). The modest evolution
of the intercept $\gamma$ and the tightness of the FP at $0.1<z<1$ can
be explained in terms of passive evolution and an old age of the
stellar populations in cluster E/S0.  The absence of a dramatic
evolution in the slopes $\alpha$ and $\beta$ argues against the
interpretation of the FP ``tilt'' resulting solely from a mass-age
relation.

So far, most of the studies have been focused on the cluster
environment. However, the population of galaxies in clusters is likely
to evolve with redshift by accretion of field galaxies. Therefore, in
order to obtain a complete and reliable picture, it is necessary to
study the evolution of E/S0 both in clusters and in the field. In
addition, hierarchical clustering models (e.~g., Kauffmann 1996)
predict the stellar populations of field E/S0 to be significantly
younger than the ones of cluster E/S0. The effects of age differences
are difficult to observe in the local Universe (Bernardi et al.\
1998), when the typical population ages are large, but are greatly
enhanced at intermediate redshift. For these reasons, we have embarked
on a campaign aimed at measuring the evolution of the FP of {\it
field} E/S0 with redshift. In previous papers (Treu et al.\ 1999;
2001a, b; hereafter T01a,b), we have analyzed a sample of 19 E/S0,
finding that a tight FP exists in the field out to $z\approx0.4$ and
that the evolution of the intercept is marginally faster than in the
clusters, indicating a marginally younger age for field galaxies (see
also van Dokkum et al.\ 2001 and Kochanek et al.\ 2000).

Here we present results from a larger sample of 30 E/S0 extended to
$z\sim0.7$, increasing dramatically our sensitivity to differences
between cluster and field environment, and show that field E/S0 evolve
faster than cluster ones at the 95\% CL. In addition, we report on the
detection of [OII]3727 emission in a significant fraction of {\it
massive} field E/S0 at intermediate redshift, which we interpret as
evidence for secondary episodes of star-formation.

We assume the Hubble constant, the matter density, and the
cosmological constant to be respectively H$_0=50h_{50}$\kms Mpc$^{-1}$
($h_{50}=1.3$ when necessary), $\Omega=0.3$, and
$\Omega_{\Lambda}=0.7$.

\section{Sample Selection and Observations}

\label{sec:data}

The sample of E/S0 was selected on the basis of Hubble Space Telescope
(HST) images taken from the HST-Medium Deep Survey (MDS; Griffiths et
al.\ 1994). The selection criteria are identical to the ones
extensively discussed in T01a, except for the color cut
($1.25<$~\Vs-\Io, chosen to select higher redshift E/S0; \Vs\, and
\Io\, indicate Vega magnitudes through HST filters F606W and F814W
respectively) and the magnitude range ($19.3<$~\Io~$<20.3$). The
effects of color and magnitude selection are taken into account in the
analysis presented here as discussed in T01b.  Images are available at
the MDS web site at http://www.archive.stsci.edu/mds. Morphological
classification was performed independently by two of us (TT, MS),
based on visual inspection of the images, of the residuals from the
fit, and of the luminosity profiles. In particular, since
contamination from Sa galaxies\footnote{See, e.g., Smail et al.\
(1997) and van Dokkum et al.\ (1998a) for a discussion.} could bias
our results towards a younger age of the stellar populations, we were
extremely conservative in rejecting galaxies that showed sign of a
spiral disk either in the direct images (F606W and F814W), or in the
residuals from the $r^{1/4}$ fit. As independent check of our
classification, the bulge-to-total luminosity ratio measured by the
MDS group for our E/S0 is always larger than 0.6 and typically in the
range $0.8$-$1$.

\begin{inlinefigure}
\begin{center}
\resizebox{\textwidth}{!}{\includegraphics{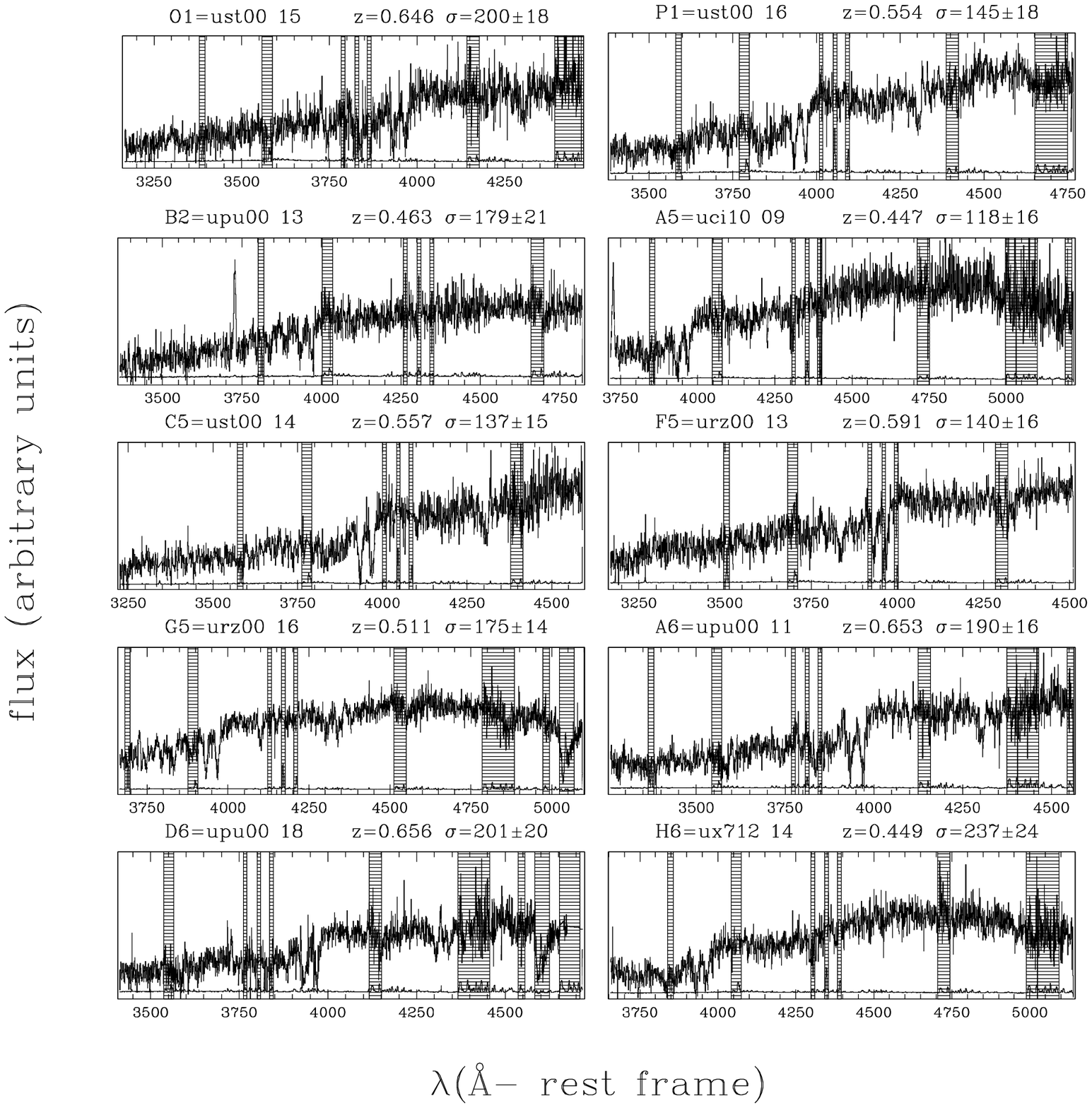}}
\end{center}
\figcaption{Spectra of galaxies with measured central velocity
dispersion. The shaded regions are affected by sky-line residuals and
atmospheric absorption and have been masked out during the fit. Note
[OII] in emission in galaxies {\bf B2} and {\bf A5}.\label{fig:spec}}
\end{inlinefigure}

Spectra for 16 galaxies were obtained in service mode from April 2000
to March 2001, using the Focal Reducer and Spectrograph 2 (FORS2) at
the Very Large Telescope (VLT). The adopted grism R600 with a $1''$
wide slit gives a resolution of $\sim 90-100$\kms; exposure times
ranged between $2\times1800$s and $2\times3600$s. For 10 out of 16
E/S0 observed (Figure~1) we obtained a reliable central velocity
dispersion (the success rate depending on observing conditions and
redshift, since the observed wavelength range was optimized for the
range $z\approx0.45-0.65$). The data reduction was very similar to the
one described in T01a. Combined with the data of T01b, our total
sample consists of 42 galaxies (in the range $z=0.10-0.73$), 30 of
which with measured velocity dispersion ($z=0.10-0.66$).

\section{The evolution of the Fundamental Plane}

Panels (a) to (e) in Figure~\ref{fig:FPB} show the location in the
FP-space of the galaxies in our sample, binned in redshift.  The data
(open symbols) in the rest-frame B band\footnote{The analysis in the V
band, or using different Coma FP, yields the same results, see T01b
for discussion.} are compared to the local relation shown as solid
line. The main result is that the brightening at fixed $\sigma$ and
\Rekpc\, increases with redshift. By assuming constant slopes $\alpha$
and $\beta$ we obtain the average evolution of the intercept with
redshift plotted as filled pentagons in panel (f).

\begin{inlinefigure}
\begin{center}
\resizebox{\textwidth}{!}{\includegraphics{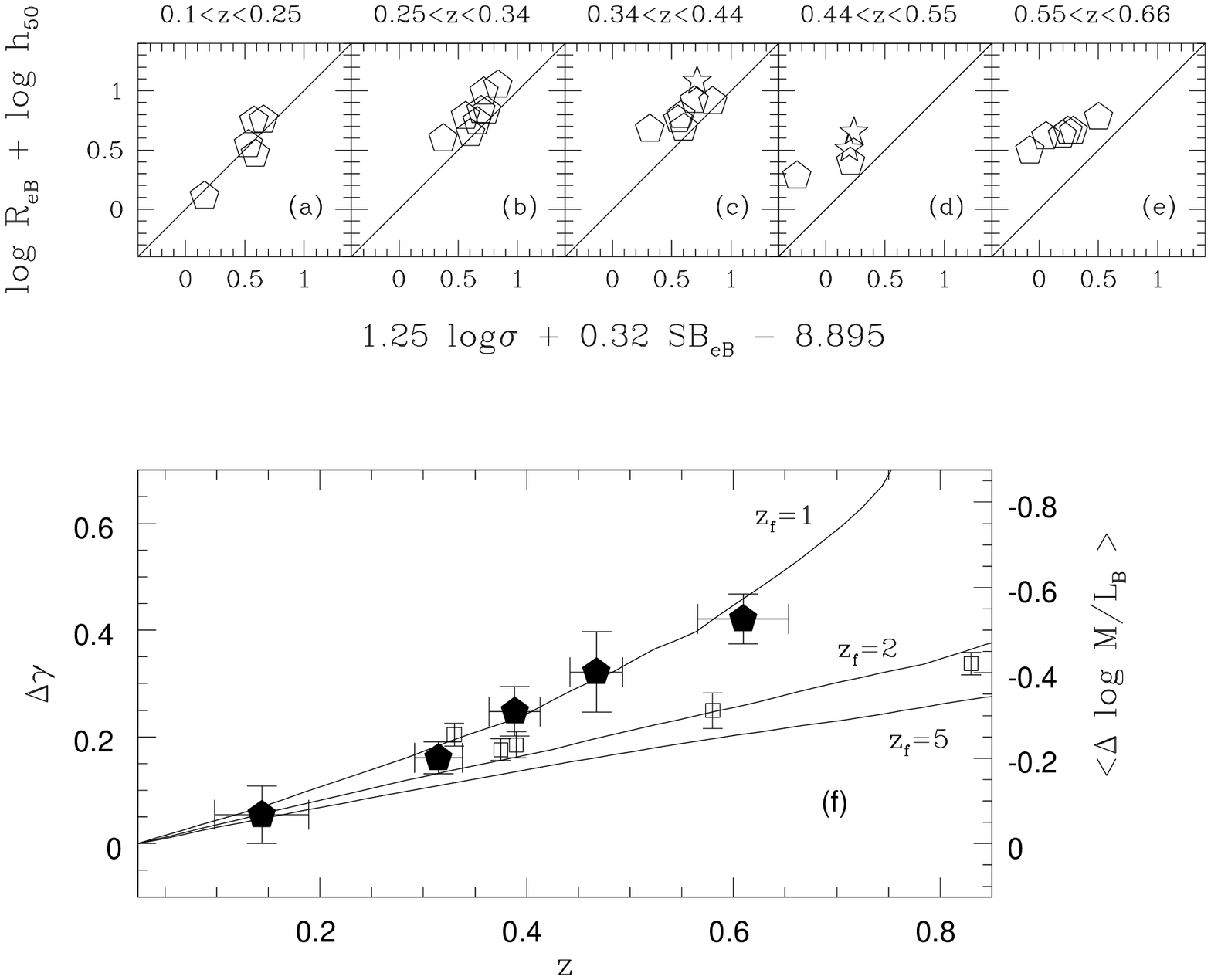}}
\end{center}
\figcaption{FP in the rest-frame B band. In panels (a) to (e) we show the
field E/S0 (open stars if [OII] emission is detected, open pentagons
otherwise), binned in redshift and compared to the FP found in the
Coma Cluster (Bender et al.\ 1998). Panel (f) shows the average offset
of the intercept of field galaxies from the local FP relation as a
function of redshift (large filled pentagons), compared to the offset
observed in clusters (open squares; van Dokkum \& Franx 1996; Kelson
et al.\ 1997; Bender et al.\ 1998; van Dokkum et al.\ 1998; Kelson et
al.\ 2000). The solid lines represent the evolution predicted for
passively evolving stellar populations formed in a single burst at
$z=1,2,5$ (from top to bottom) computed using Bruzual \& Charlot
(1993) models in the BC96 version (as described in T01b). We assume
$\Omega=0.3$, $\Omega_{\Lambda}=0.7$ and h$_{50}=1.3$
\label{fig:FPB}}
\end{inlinefigure}

Under the assumption of passive evolution and constant slopes (T01b),
the evolution of $\gamma$ can be converted into the evolution of the
average $M/L$ ratio, which can be read on the right scale of panel
(f). Modelling the evolution as $\gamma(z)=\gamma(0)+\gamma'(0) z$,
where the prime indicates derivative with respect to $z$, a least
$\chi^2$ fit yields $\gamma'_{B}=0.64$. However, in order to avoid
biased results, it is crucial to take into account the selection
process, for example by applying the Montecarlo-Bayesian method
introduced in T01b, generalized by allowing the scatter to vary as a
function of redshift,
$\sigma_{\gamma}(z)=\sigma_{\gamma}(0)+\sigma_{\gamma}'(0)z$. In this
way, we derive the posterior probability density given the set of
observations, p($\gamma',\sigma_{\gamma}' | \{\gamma_i\}$) in the
notation of T01b, shown in Figure~\ref{fig:prob} panel (a). The
posterior probability peaks at $\gamma'_B=0.58$ ($0.45-0.67$ 68\%
limit), which corresponds to $\log
(M/L_B)'=-0.72^{+0.11}_{-0.16}$. The scatter is constant or at most
mildly increasing with redshift. Note that the effects of the
magnitude selection limit (see T01b) on the measured evolution of the
intercept depend on the intrinsic scatter of the FP, and therefore it
is important to study the combined probability density.  Note also
that the galaxies with OII emission (open stars in panels c and d)
fall slightly above the quiescent ones (pentagons), as expected if
their M/L is smaller because of a fraction of young stars. However,
they do not dominate the general trend, since the results change very
little if they are removed from the sample (e.g. $\gamma'_B$ from the
least $\chi^2$ fit goes from 0.64 to 0.61).
\begin{inlinefigure}
\begin{center}
\resizebox{\textwidth}{!}{\includegraphics{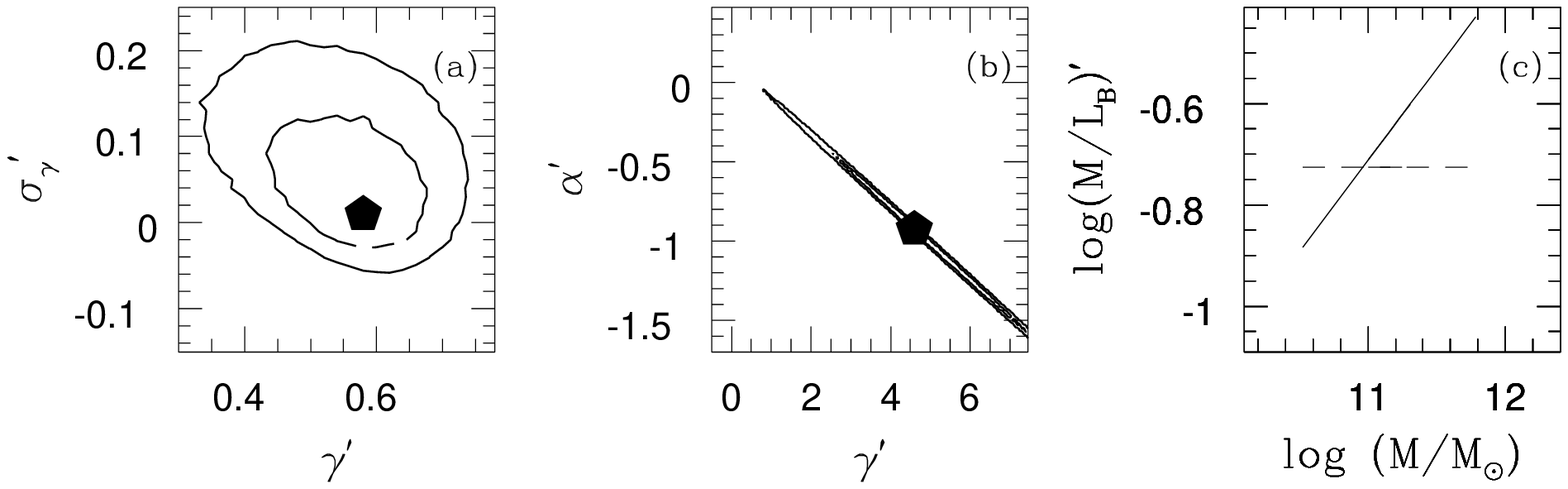}}
\end{center}
\figcaption{Panel (a): probability contours for the variation of the
intercept and scatter with redshift \label{fig:prob} (respectively
$\gamma'$ and $\sigma'_{\gamma}$; the pentagon marks the peak of the
probability density, the dashed line the 68\% limit and the solid line
the 95\% limit). Panel (b): probability contours for the variation of
the intercept ($\gamma'$) and slope ($\alpha'$) with redshift,
obtained by assuming $\alpha'=10\beta'$; symbols as in panel
(a). Panel (c): evolution of $M/L_B$ as a function of M (defined as
$5\sigma^2R_e/G$), as obtained by assuming fixed slopes (dashed line),
or the values of $\alpha'$ and $\beta'$ corresponding to the peak of
the probability density in panel (b; solid line).}
\end{inlinefigure}

It is desirable to generalize the treatment by allowing also the
slopes to evolve with redshift. Even though a larger sample is needed
to this aim, possibly spanning a wider range of parameters, we will
try to address this issue by assuming two additional constraints: {\it
i)} the scatter does not change from $z=0$ to $z\sim0.7$ and {\it ii)
}$\alpha'=10\beta'$ (hence the FP is equivalent to a relation between
M and M/L at any redshift).  Under these assumptions, we obtain the
results shown in Fig.~\ref{fig:prob} panels (b) and (c), plotted using
$M=5\sigma^2R_e/G$ (Bender, Burstein \& Faber 1992). The evolution of
the slopes remains highly unconstrained (panel b), but the evolution
of the $M$ vs $M/L$ relation is better constrained (panel c). For the
typical masses in our sample ($<\log M/M_{\odot}>=11.4$) the derived
value of $M/L'$ does not depend significantly on the value of
$\alpha'$, providing a test of the robustness of the measurement. On
the other hand, the result shown as a solid line in panel (c) would
indicate that the evolution of $M/L$ with redshift becomes slower as
the mass increases (i.e. for example massive E/S0 are older). More
data are needed to confirm these findings.

In Figure~\ref{fig:FPB} we also show the evolution of $\gamma$
predicted by assuming passively evolving single stellar
populations. The best-fitting single burst redshift of formation
should be interpreted in terms of a luminosity weighted redshift of
formation, since it is clear that a more complex scenario than passive
evolution of single coeval stellar populations is needed to explain
the present observations (T01b). Finally, cluster data taken from the
literature are shown as open squares in Figure~\ref{fig:FPB} for
comparison. The field FP evolves faster than the cluster one: for the
cluster van Dokkum et al.\ (2001) measure $\log
(M/L_B)'=-0.49\pm0.05$, i.~e. $\gamma'=0.39\pm0.04$, which falls on
the 95\% contour in panel (a) of Figure~\ref{fig:prob}.

\section{Emission line properties}

At $z\ga0.3$ the emission line [OII]3727 is redshifted into the
wavelength range covered by our instrumental setup. Six out of 27 E/S0
for which [OII] is detectable show significant emission (rest-frame
equivalent width EW$>5$ \AA). If attributed to star-formation, the
observed [OII] EW correspond to star-formation rates of order
$0.5$--$5$ $M_{\odot}/yr$ (Kennicutt 1992). The fraction of E/S0 with
sizable star-formation in our sample is significantly larger than the
one found for massive E/S0 in clusters at similar redshifts and in the
local Universe. In Figure~\ref{fig:OII} we plot the fraction of E/S0
with [OII] EW$>5$ \AA\, as a function of redshift for field
(pentagons) and cluster (squares) samples taken from the literature.

\begin{inlinefigure}
\begin{center}
\resizebox{\textwidth}{!}{\includegraphics{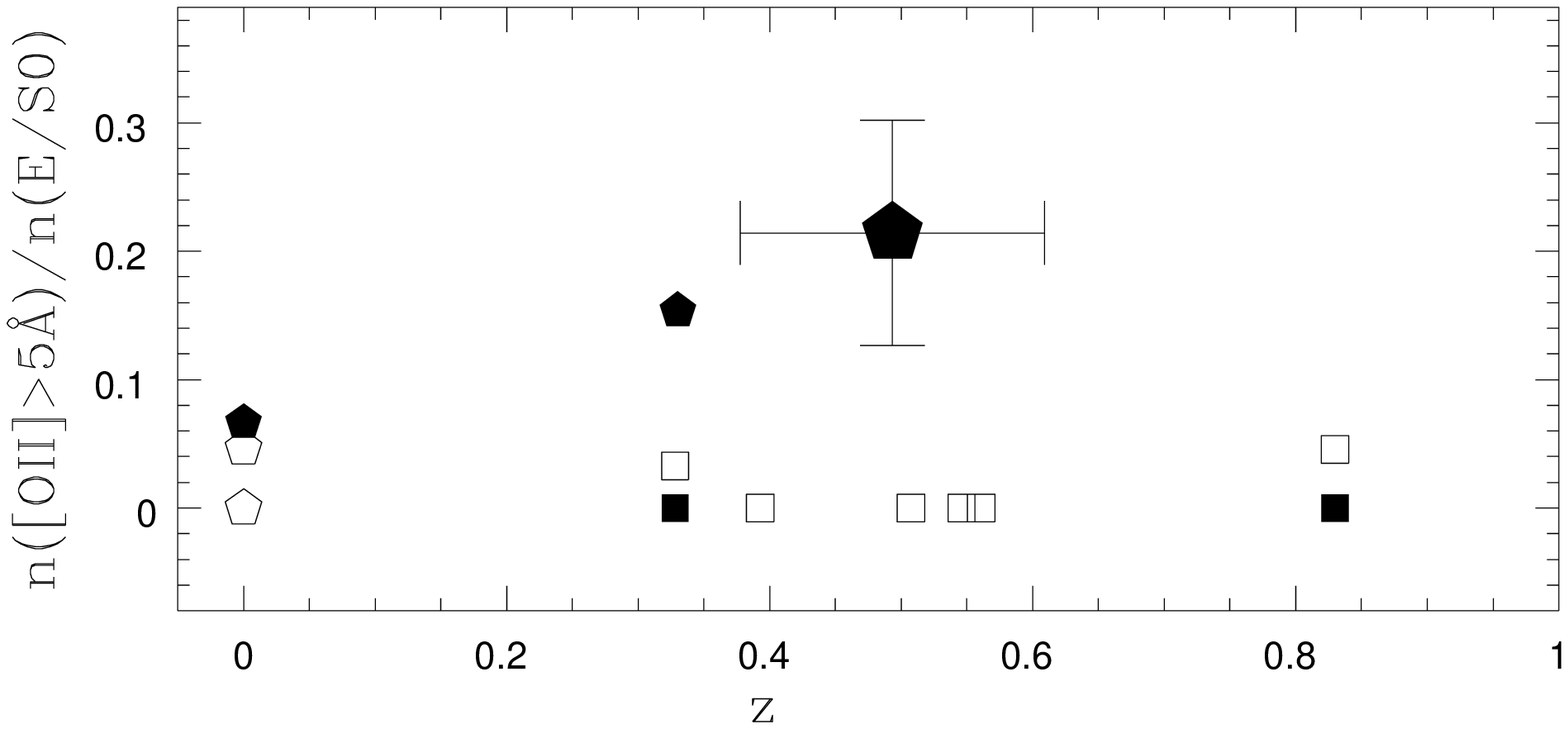}}
\end{center}
\figcaption{Fraction of E/S0 with [OII] emission larger than 5 \AA\,
EW (rest-frame) as a function of redshift. Pentagons represent field
galaxies, squares cluster galaxies. The points are filled if selected
similarly to our sample.  Field galaxies are from Caldwell (1984),
Kennicutt (1992), Jansen et al.\ (2000; limited to M$_B-5\log
h_{50}<-20$, corresponding roughly to our survey), Brinchmann et al.\
(1998; limited to I$_8<20.3$), and the present work (large filled
pentagon with error bars).  The filled squares are from Fisher et al.\
(1998), van Dokkum et al.\ (1998a), van Dokkum et al.\ (2000). The
open squares are from Smail et al.\ (1997) and Dressler et al.\
(1999). \label{fig:OII} }
\end{inlinefigure}

E/S0 with similar [OII] EW, or similarly with relatively blue colors,
have been reported previously, suggesting that secondary episodes of
star-formation are common at intermediate redshift (Schade et al.\
1999; Menanteau et al.\ 2001). Im et al.\ (2001) measured the width of
the [OII] line for a sample of blue E/S0 at intermediate redshift
finding velocity dispersions $\sigma\la80$ kms$^{-1}$; based on this
measurement they argue that most blue E/S0 are not the progenitors of
present-day massive E/S0, but rather less massive spheroids, for which
star-formation is observed also in the local Universe (Jansen et al.\
2000). Our sample adds further information, because it is made of
bright ($I<20.3$) and {\it massive} E/S0. The velocity dispersions we
obtained via absorption-line kinematics are typical of massive field
E/S0. In particular, the three objects with significant [OII] emission
for which velocity dispersion is available have $\sigma=118, 179, 233$
km s$^{-1}$. From the FP relationship, as measured from our sample, we
estimate for the other three galaxies with [OII] emission $\sigma=105,
171, 261$ kms$^{-1}$.

We can estimate the amount of stellar mass assembled in these
secondary burst in the following way. The probability of observing a
burst is given by
\begin{equation}
p(\tx{OII})={\bar n}\frac{{\bar t}}{\Delta t},
\label{eq:OII1}
\end{equation}
where ${\bar n}$ is the average number of bursts per galaxy, ${\bar
t}$ is the average duration of the burst, and $\Delta t$ is the
interval in cosmic time between $z=0.73$ and $z=0.26$ when the burst
was observable. The total average mass in secondary bursts is then:
\begin{equation}
M_{\tx{OII}}={\bar n}{\bar t}{\dot {\bar M}}=p(\tx{OII}) \Delta t
{\dot {\bar M}}.
\label{eq:OII2}
\end{equation}
By using the observed values for $p(\tx{OII})=6/27$ and ${\dot {\bar
M}}=0.5-5M_{\odot}$yr$^{-1}$, we estimate that the average
stellar mass formed in these bursts is of the order of
$M_{\tx{OII}}\sim 5\cdot10^{8} - 5\cdot10^{9} M_{\odot}$.

\section{Conclusions}

Our sample of 30 field E/S0 defines a tight FP out to $z=0.66$, with
no or modest increase of the scatter with redshift. The intercept
$\gamma$ evolves with redshift as $\gamma'=0.58$, which, interpreted
in terms of passive evolution of the stellar populations, implies
$\log(M/L_B)'=-0.72^{+0.11}_{-0.16}$, i.e. faster than that observed in
clusters by other groups (e.g. $-0.49\pm0.05$ van Dokkum et al.\
2001). In addition, 22\% of the galaxies show [OII] in emission, with
$EW>5$\AA. Assuming the emission is due to star-formation, we estimate
that such bursts produce of order $M_{\tx{OII}}\sim 5\cdot10^{8} -
5\cdot10^{9} M_{\odot}$ of stellar mass between $z=0.73$ and $z=0.26$.

The existence and tightness of the FP suggest that no major structural
changes occur between $z\sim0.7$ and $z=0$.  However, we find evidence
that some stellar mass is formed at relatively recent time during
secondary bursts. Although these bursts contribute only a small
fraction of the total stellar mass they contribute significantly to
the evolution of the observable luminous component. For example, a
scenario where most of the stars formed at $z>1$ and secondary
star-formation occurs at $z<1$ could not only explain the evolution of
the FP (T01b) and the observed [OII] emission but could also reconcile
the modest evolution of the number density of E/S0 observed to
$z\sim1$ (Schade et al.\ 1999; Im et al.\ 2001, in press), with the
paucity of {\it red} E/S0 observed in some infrared surveys (Menanteau
et al.\ 1999; Treu \& Stiavelli 1999; see also Jimenez et al.\ 1999
and Daddi et al.\ 2000).

\acknowledgments

This research was supported by STScI grant AR-09222.  We acknowledge
the use of the Gauss-Hermite Fourier Fitting Software developed by
R.~van der Marel and M.~Franx and useful conversations with A.~Benson,
R.~Ellis, F.~Menanteau, and P.~van Dokkum.

\end{document}